\documentclass[12pt]{article}%
\usepackage{amsmath}%
\usepackage{amsfonts}%
\usepackage{amssymb}%
\usepackage{graphicx}

\begin{document}

\title{Recovering the equivalence of ensembles II: An Ising chain with competing
short and long-range interactions}
\author{Vera B. Henriques and S. R. Salinas\\Instituto de F\'{\i}sica,\\Universidade de S\~{a}o Paulo,\\S\~{a}o Paulo, SP, Brazil}
\date{August 14, 2015}
\maketitle

\begin{abstract}
In a pioneer work, John Nagle has shown that an Ising chain with competing
short and long-range interactions displays second and first-order phase
transitions separated by a tricritical point. More recently, it has been
claimed that Nagle%
\'{}%
s model provides an example of the inequivalence between canonical and
microcanonical calculations. We then revisit Nagle%
\'{}%
s original solution, as well as the usual formulation of the problem in a
canonical ensemble, which lead to the same results. Also, in contrast to
recent claims, we show that an alternative formulation in the microcanonical
ensemble, with the adequate choice of the fixed thermodynamic extensive
variables, leads to equivalent thermodynamic results.

\end{abstract}

\section{Introduction}

In the beginning of the seventies, John Nagle \cite{nagle1970}%
\cite{bonner1971} analyzed an Ising chain with antiferromagnetic interactions
between nearest-neighbor sites, and the addition of equivalent-neighbor
(mean-field) ferromagnetic interactions between all pairs of sites. Depending
on the strength of the competition, this system was shown to display second
and first-order phase transitions separated by a \textquotedblleft special
critical point\textquotedblright, which was later named a tricritical point
\ \cite{griffiths1972}. A few years ago, this problem has been revisited by
some authors as one on the \textquotedblleft paradigmatic
examples\textquotedblright\ of the inequivalence of ensembles, in which the
very localization of the tricritical point was supposed to depend on the
ensemble (canonical or microcanonical) that was used to carry out the
statistical calculations \cite{mukamel2005}\cite{campa2009}. We have recently
disproved similar claims of inequivalence of ensembles for a long-range
version of a spin-$1$ Ising model \cite{henriques2015}. In the present article
we give arguments to show the equivalence of solutions in Nagle%
\'{}%
s model.

The Hamiltonian of Nagle%
\'{}%
s model in zero external field may be written as%
\begin{equation}
\mathcal{H}=-J_{SR}\sum_{i=1}^{N}\sigma_{i}\sigma_{i+1}-\frac{1}{2N}%
J_{LR}\left(  \sum_{i=1}^{N}\sigma_{i}\right)  ^{2},\label{HN}%
\end{equation}
where $\sigma_{i}=\pm1$ for $i=1$, $2$, $...$, $N$, the long-range
interactions are ferromagnetic, $J_{LR}>0$, and the presence of a tricritical
point requires antiferromagnetic short-range interactions ($J_{SR}<0$). In his
original work, Nagle obtained exact thermodynamic solutions by two elegant and
complementary techniques, which already refer to different thermodynamic
representations. Later, this model was solved by easier manipulations, in the
usual canonical ensemble \cite{kardar1983}\cite{kislinsky1988}%
\cite{vieira1995}. Due to its instructive features, and to a number of
misconceptions in the literature, we begin by reviewing Nagle%
\'{}%
s solution. We then resort to a Gaussian identity to establish the (same)
solutions in the usual canonical ensemble. Finally, we use the corresponding
Ising chain, with the exclusion of the long-range terms, to write an entropy
function in the microcanonical ensemble. With the appropriate choice of the
extensive variables, and properly accounting for the long-range interactions,
we show that there are no discrepancies between canonical and microcanonical results.

\section{Original solutions of Nagle}

In the more detailed solution of the problem, Nagle considers the Hamiltonian
of an Ising chain with the exclusion of the long-range interactions
($J_{LR}=0$) and in the presence of a field $H$,
\begin{equation}
\mathcal{H}_{I}=-J_{SR}\sum_{i=1}^{N}\sigma_{i}\sigma_{i+1}-H\sum_{i=1}%
^{N}\sigma_{i}.
\end{equation}
Given the temperature $k_{B}T=1/\beta$ and the field $H$, we write the usual
form of the canonical partition function,%
\begin{equation}
Z_{I}=Z_{I}\left(  T,H,N\right)  =%
{\displaystyle\sum_{\left\{  \sigma_{i}\right\}  }}
\exp\left[  \beta J_{SR}\sum_{i=1}^{N}\sigma_{i}\sigma_{i+1}+\beta H\sum
_{i=1}^{N}\sigma_{i}\right]  ,
\end{equation}
which can be analytically obtained by the transfer matrix technique. In the
thermodynamic limit, the associated free energy per site, as a function of $T
$ and $H$, is given by%
\begin{equation}
g_{I}=g_{I}\left(  T,H\right)  \sim-\frac{1}{\beta N}\ln Z_{I},
\end{equation}
from which we obtain the magnetization per site,%
\begin{equation}
m=\frac{1}{N}\left\langle \sum_{i=1}^{N}\sigma_{i}\right\rangle =-\left(
\frac{\partial g_{I}}{\partial H}\right)  _{T}.\label{mag}%
\end{equation}
We then use a Legendre transformation to write another free energy,
$f_{I}=f_{I}\left(  T,m\right)  $, which is expressed as a function of
temperature $T$ and magnetization $m$,
\begin{equation}
f_{I}\left(  T,m\right)  =g_{I}\left(  T,H\right)  +mH.
\end{equation}
In this one dimensional system, there are no problems of convexity, and the
field $H$ can be eliminated by using equation (\ref{mag}).

Nagle then remarks that the energy per site of an additional term, of
mean-field nature, may be written in terms of the magnetization, so that we
have%
\begin{equation}
\frac{1}{N}\left\langle -\frac{1}{2N}J_{LR}\left(  \sum_{i=1}^{N}\sigma
_{i}\right)  ^{2}\right\rangle =-\frac{1}{2}J_{LR}m^{2}.
\end{equation}
Taking into account that this term depends only on $m$, the free energy
$f=f\left(  T,m\right)  $, associated with Nagle%
\'{}%
s model, defined by the Hamiltonian of equation (\ref{HN}), is given by%
\begin{equation}
f=f\left(  T,m\right)  =f_{I}\left(  T,m\right)  -\frac{1}{2}J_{LR}m^{2}.
\end{equation}
This is the central equation of Nagle%
\'{}%
s treatment. In analogy with a Landau expansion, the free energy $f\left(
T,m\right)  $ may be written as a power series in terms of the magnetization,%
\begin{equation}
f\left(  T,m\right)  =a_{0}\left(  T\right)  +a_{2}\left(  T\right)
m^{2}+a_{4}\left(  T\right)  m^{4}+a_{6}\left(  T\right)  m^{6}%
+...,\label{ftm}%
\end{equation}
from which we obtain the critical line ($a_{2}=0$; $a_{4}>0$) and the location
of the tricritical point ($a_{2}=a_{4}=0$; $a_{6}>0$). It should be noted that
this Landau expansion is written in terms of a density (the order parameter
$m$) and that the coefficients of this expansion depend on the thermodynamic
fields (in this case, the parameters $\beta J_{SR}$ and $\beta J_{LR}$).

In the Appendix of his article, Nagle mentions an alternative calculation, in
which the canonical partition function, in zero field, is written as%
\begin{equation}
Z\left(  T,H=0,N\right)  =%
{\displaystyle\sum\limits_{M=-N}^{N}}
\exp\left[  \beta J_{LR}\frac{M^{2}}{2N}\right]
{\displaystyle\sum\limits_{S=0}^{R}}
f_{N}\left(  R,S\right)  \exp\left[  -\beta J_{SR}\left(  N-4S\right)
\right]  ,
\end{equation}
where $R=\min\left\{  \left(  N\pm M\right)  /2\right\}  $, $S$ is the number
of $\left(  +,-\right)  $ pairs, and%
\begin{equation}
f_{N}\left(  R,S\right)  =\frac{N}{S}\left(
\begin{array}
[c]{c}%
R-1\\
S-1
\end{array}
\right)  \left(
\begin{array}
[c]{c}%
N-R-1\\
S-1
\end{array}
\right)  .
\end{equation}
Although referring to a future publication, Nagle and Yeo never published the
combinatorial derivation of $f_{N}\left(  R,S\right)  $, which corresponds to
the number of microstates of the system with fixed values of $N$, $M$ and $S$
(in other words, with fixed magnetization $m$ and internal energy $u$
associated with the short-range terms). This expression of $f_{N}\left(
R,S\right)  $ is directly related to the entropy in the microcanonical
ensemble in terms of the appropriate densities. In the thermodynamic limit,
the partition function $Z\left(  T,H=0,N\right)  $ is given by the maximum
term of the sum over $M$ and $S$. According to Nagle, all the results in zero
field have been checked in this alternative formulation, in particular the
location of the tricritical point.

The choice of independent variables and the alternative solutions of Nagle are
already a firm indication of the equivalence of ensembles. The asymptotic form
of the expression of $f_{N}\left(  R,S\right)  $, which is directly related to
the entropy in the microcanonical ensemble, has been independently written by
several authors \cite{rehn2012}, even in recent work with claims of
inequivalence of ensembles \cite{mukamel2005}. The most remarkable deduction
has been published by Ernst Ising \cite{ising1925} in his famous article of 1925.

\section{Solution in the canonical ensemble}

The usual canonical partition function associated with Hamiltonian (\ref{HN})
is given by%
\begin{equation}
Z=Z\left(  \beta J_{SR},\beta J_{LR}\right)  =%
{\displaystyle\sum\limits_{\left\{  \sigma_{i}\right\}  }}
\exp\left[  \beta J_{SR}\sum_{i=1}^{N}\sigma_{i}\sigma_{i+1}+\frac{\beta
J_{LR}}{2N}\left(  \sum_{i=1}^{N}\sigma_{i}\right)  ^{2}\right]  .
\end{equation}
Using the Gaussian identity%
\begin{equation}
\int_{-\infty}^{+\infty}dx\,\exp\left[  -x^{2}+2ax\right]  =\sqrt{\pi}%
\,\exp\left(  a^{2}\right)  ,
\end{equation}
we have%
\begin{equation}
Z=\left(  \frac{\beta J_{LR}N}{2\pi}\right)  ^{1/2}\int_{-\infty}^{+\infty
}dy\,\exp\left[  -\beta Nf\left(  y\right)  \right]  ,\label{Zcan}%
\end{equation}
where%
\begin{equation}
f\left(  y\right)  =\frac{1}{2}J_{LR}\,y^{2}-\frac{1}{\beta N}\ln Z_{I},
\end{equation}
and $Z_{I}$ is the canonical partition function of an Ising chain,%
\begin{equation}
Z_{I}=Z_{I}\left(  \beta J_{SR},\beta J_{LR}y\right)  =%
{\displaystyle\sum\limits_{\left\{  \sigma_{i}\right\}  }}
\exp\left[  \beta J_{SR}\sum_{i=1}^{N}\sigma_{i}\sigma_{i+1}+\beta J_{LR}%
y\sum_{i=1}^{N}\sigma_{i}\right]  .
\end{equation}
Also, we remark that these results can be obtained from an application of a
well-known Bogoliubov identity \cite{kislinsky1988}.

In the thermodynamic limit we write%
\begin{equation}
f\left(  y\right)  \sim\frac{1}{2}J_{LR}\,y^{2}-\frac{1}{\beta}\ln
\lambda\left(  y\right)  ,\label{fy}%
\end{equation}
where $\lambda\left(  y\right)  $ is the largest eigenvalue of a transfer
matrix,%
\begin{equation}
\lambda=\exp\left(  \beta J_{SR}\right)  \cosh\left(  \beta J_{LR}y\right)  +
\left[  \exp\left(  2\beta J_{SR}\right)  \cosh^{2}\left(  \beta
J_{LR}y\right)  -2\sinh\left(  2\beta J_{SR}\right)  \right]  ^{1/2}.
\end{equation}
We can analyze the critical behavior from an expansion of the asymptotic form
of $f\left(  y\right)  $ as a power series in $y$,%
\begin{equation}
f\left(  y\right)  =A_{0}+A_{2}\,y^{2}+A_{4}\,y^{4}+A_{6}\,y^{6}+...,
\end{equation}
which is equivalent to Nagle%
\'{}%
s expansion of the free energy $f\left(  T,m\right)  $, given by equation
(\ref{ftm}). The critical line comes from $A_{2}=0$, with $A_{4}>0$, and the
tricritical point is located at $A_{2}=A_{4}=0$, with $A_{6}>0$.

If we use Laplace%
\'{}%
s method to calculate the asymptotic form of the integral (\ref{Zcan}), the
saddle-point equation is given by%
\begin{equation}
\widetilde{y}=\frac{\sinh\left(  \beta J_{LR}\widetilde{y}\right)  \left[
1+D^{-1/2}\cosh\left(  \beta J_{LR}\widetilde{y}\right)  \right]  }%
{\cosh\left(  \beta J_{LR}\widetilde{y}\right)  +D^{1/2}},\label{eqstate}%
\end{equation}
where%
\begin{equation}
D=\sinh^{2}\left(  \beta J_{LR}\widetilde{y}\right)  +\exp\left(  -4\beta
J_{SR}\right)  ,
\end{equation}
so we have the corresponding free energy per spin,%
\begin{equation}
g=g\left(  T\right)  =-\frac{1}{2}J_{LR}\widetilde{y}^{2}-\frac{1}{\beta}%
\ln\lambda\left(  \widetilde{y}\right)  .
\end{equation}
In the next Section we derive again the equation of state (\ref{eqstate}) in
the context of the microcanonical formulation. As in a typical mean-field
calculation, there is always a paramagnetic solution, $\widetilde{y}=0$, but
this solution becomes physically unacceptable in the ordered region of the
phase diagram. If there are several solutions, we have to choose the absolute
minima, which corresponds to using a Maxwell construction (and to recovering
the convexity of the free energy). From the equation of state (\ref{eqstate}),
it is possible to check the location of the tricritical point, given by%
\begin{equation}
\beta J_{LR}=\exp\left(  -2\beta J_{SR}\right)  ,
\end{equation}
which corresponds to $A_{2}=0$, and
\begin{equation}
\beta J_{LR}\left[  \frac{1}{3}+\frac{4}{3}\exp\left(  2\beta J_{SR}\right)
-\exp\left(  6\beta J_{SR}\right)  \right]  =1+\exp\left(  2\beta
J_{SR}\right)  ,
\end{equation}
which corresponds to $A_{4}=0$, in agreement with Nagle%
\'{}%
s findings.

\section{Solution in the microcanonical ensemble}

According to the work of Nagle, it is convenient to begin by considering the
Hamiltonian of an Ising chain, without the addition of mean-field terms and in
the presence of an external field $H$, which can be written as%
\begin{equation}
\mathcal{H}_{I}=-J_{SR}\sum_{i=1}^{N}\sigma_{i}\sigma_{i+1}-H\sum_{i=1}%
^{N}\sigma_{i}=U-HM,
\end{equation}
where the energy $U$ refers to the short-range interactions,%
\begin{equation}
U=-J_{SR}\sum_{i=1}^{N}\sigma_{i}\sigma_{i+1};\qquad M=\sum_{i=1}^{N}%
\sigma_{i}.
\end{equation}

Given $U$, $M$, and $N$, the number of microscopic states associated with this
system may be formally written as a sum over spin configurations of a product
of two delta functions,%
\begin{equation}
\Omega_{I}=\Omega_{I}\left(  U,M,N\right)  =\sum_{\left\{  \sigma_{i}\right\}
}\delta\left(  U+J_{SR}\sum_{i=1}^{N}\sigma_{i}\sigma_{i+1}\right)
\,\delta\left(  M-\sum_{i=1}^{N}\sigma_{i}\right)  .
\end{equation}
We now introduce integral representations of these delta functions, and use
the transfer matrix technique to carry out the sums. In the thermodynamic
limit, it is straightforward to write%
\begin{equation}
\Omega_{I}\left(  U,M,N\right)  \sim\int\int dk_{1}dk_{2}\,\exp\left[
N\,f\left(  k_{1},k_{2}\right)  \right]  ,
\end{equation}
where%
\begin{equation}
f\left(  k_{1},k_{2}\right)  =k_{1}u-k_{2}m+\ln\lambda\left(  k_{1}%
,k_{2}\right)  ,
\end{equation}
and%
\begin{equation}
\lambda\left(  k_{1},k_{2}\right)  =\exp\left(  k_{1}J_{SR}\right)  \left\{
\cosh k_{2}+\left[  \sinh^{2}k_{2}+\exp\left(  -4k_{1}J_{SR}\right)  \right]
\right\}  ,
\end{equation}
with $u=U/N$ and $m=M/N$.

The entropy per particle as a function of $u$ and $m$ is given by%
\begin{equation}
s_{I}=s_{I}\left(  u,m\right)  \sim\frac{k_{B}}{N}\ln\Omega_{I}\sim
k_{B}\,f\left(  \widetilde{k}_{1},\widetilde{k}_{2}\right)  ,\label{sI}%
\end{equation}
where $\widetilde{k}_{1}$ and $\widetilde{k}_{2}$ come form the saddle-point
equations, $\left(  \partial f/\partial k_{1}\right)  _{k_{2}}=0$ and $\left(
\partial f/\partial k_{2}\right)  _{k_{1}}=0$,
\begin{equation}
u+J_{SR}=2J_{SR}\frac{\exp\left(  -4\widetilde{k}_{1}\right)  D^{-1/2}}%
{\cosh\widetilde{k}_{2}+D^{1/2}}\label{umicro}%
\end{equation}
and%
\begin{equation}
m=\frac{\sinh k_{2}\left[  1+D^{-1/2}\cosh k_{2}\right]  }{\cosh\widetilde
{k}_{2}+D^{1/2}},\label{mmicro}%
\end{equation}
with%
\begin{equation}
D=\sinh^{2}k_{2}+\exp\left(  -4k_{1}J_{SR}\right)  .\label{Dmicro}%
\end{equation}

In the entropy representation, we write the differential form%
\begin{equation}
ds_{I}=\frac{1}{T}du-\frac{H}{T}dm,
\end{equation}
from which we have the equations of state,%
\begin{equation}
\frac{1}{T}=\left(  \frac{\partial s_{I}}{\partial u}\right)  _{m}%
;\qquad-\frac{H}{T}=\left(  \frac{\partial s_{I}}{\partial m}\right)  _{u}.
\end{equation}
It is straightforward to use these equations, together with the saddle point
equations (\ref{umicro}) and (\ref{mmicro}), in order to show that%
\begin{equation}
\widetilde{k}_{2}=\beta H;\qquad\widetilde{k}_{1}=\beta,\label{k1k2}%
\end{equation}
which is an evidence of the equivalence of ensembles (in the absence of the
mean-field interactions).

We now turn to Nagle%
\'{}%
s model, with the addition of the long-range terms. In the presence of the
equivalent-neighbor interactions, the internal energy is given by the sum of
two terms,
\begin{equation}
u=u_{SR}+u_{LR}=u_{SR}-\frac{1}{2}J_{LR}\,m^{2},\label{u}%
\end{equation}
so that both the energy associated with the short-range interactions, $u_{SR}
$, and the magnetization $m$ should be fixed in the microcanonical
formulation. Therefore, the entropy of Nagle%
\'{}%
s model is still given by the expression $s_{I}$, as in equation (\ref{sI}),
but with the energy given by equation (\ref{u}), which leads to a new
differential form,%
\begin{equation}
ds=\frac{1}{T}du_{SR}-\frac{J_{LR}\,m}{T}dm-\frac{H}{T}dm.
\end{equation}
From this expression we have%
\begin{equation}
\frac{1}{T}=\left(  \frac{\partial s}{\partial u_{SR}}\right)  _{m}%
;\qquad-\frac{J_{LR}\,m}{T}-\frac{H}{T}=\left(  \frac{\partial s}{\partial
m}\right)  _{u_{SR}},
\end{equation}
where $H_{ef}=H+J_{LR}\,m$ is an effective field, including the external field
$H$ and the effects of the long-range terms. Thus, in zero external field, we
use equations (\ref{k1k2}) to write $\widetilde{k}_{2}=\beta J_{LR}\,m$ and
$\widetilde{k}_{1}=\beta$. Inserting these expressions into equations
(\ref{mmicro}) and (\ref{Dmicro}), we obtain%
\begin{equation}
m=\frac{\sinh\left(  \beta J_{LR}m\right)  \left[  1+D^{-1/2}\cosh\left(
\beta J_{LR}m\right)  k_{2}\right]  }{\cosh\left(  \beta J_{LR}m\right)
+D^{1/2}},
\end{equation}
where%
\begin{equation}
D=\sinh^{2}\left(  \beta J_{LR}m\right)  +\exp\left(  -4\beta J_{SR}\right)  ,
\end{equation}
which is identical to the equation of state (\ref{eqstate}) in the canonical
ensemble, with the identification of $m$ with $\widetilde{y}$ (and which
already leads to the location of the tricritical point). In contrast to
previous calculations, we do not find any disagreements in the thermodynamic
behavior obtained from calculations in different ensembles.

\section{Conclusions}

We revisited the statistical analysis of a spin-$1/2$ Ising chain with
antiferromagnetic interactions between nearest-neighbor sites, and the
addition of equivalent-neighbor ferromagnetic interactions between all pairs
of sites. This system, which is known to display second and first-order phase
transitions separated by a tricritical point, has been used as one of the
paradigmatic examples of inequivalence of canonical and microcanonical
formulations. In contrast to these claims, we give arguments to show the
equivalence of thermodynamic solutions in different ensembles.

\end{document}